\documentclass{article}
\usepackage{spconf}

\usepackage{cite,hyperref,balance,color}
\usepackage{amsmath,amssymb,amsfonts}
\usepackage{algorithmic}
\usepackage{graphicx,booktabs,multirow}
\usepackage{textcomp}
\usepackage{xcolor}
\usepackage{tcolorbox}
\usepackage{adjustbox}
\tcbset{width=0.9\textwidth,boxrule=0pt,colback=red,arc=0pt,auto outer arc,left=0pt,right=0pt,boxsep=5pt}

\begin{document}
\ninept

\title{
Leveraging Pretrained Representations with Task-\\
related Keywords for Alzheimer's Disease Detection
}

\name{
Jinchao Li$^1$, Kaitao Song$^2$, Junan Li$^1$, Bo Zheng$^1$, Dongsheng Li$^2$, Xixin Wu$^1$, Xunying Liu$^1$, Helen Meng$^1$
\thanks{$^1$The work was done during the author’s internship with Microsoft.}
}
\address{
$^1$The Chinese University of Hong Kong, Hong Kong SAR, China\\
$^2$Microsoft Research Asia, Shanghai, China\\
{\tt \small $^1$\{jcli,jli,bzheng,wuxx,xyliu,hmmeng\}@se.cuhk.edu.hk, $^2$\{kaitaosong,Dongsheng.Li\}@microsoft.com}
}

\maketitle

\begin{abstract}
With the global population aging rapidly, Alzheimer’s disease (AD) is particularly prominent in older adults, which has an insidious onset and leads to a gradual, irreversible deterioration in cognitive domains (memory, communication, etc.). Speech-based AD detection opens up the possibility of widespread screening and timely disease intervention.
Recent advances in pre-trained models motivate AD detection modeling to shift from low-level features to high-level representations.
This paper presents several efficient methods to extract better AD-related cues from high-level acoustic and linguistic features.
Based on these features, the paper also proposes a novel task-oriented approach by modeling the relationship between the participants' description and the cognitive task.
Experiments are carried out on the ADReSS dataset in a binary classification setup, and models are evaluated on the unseen test set. Results and comparison with recent literature demonstrate the efficiency and superior performance of proposed acoustic, linguistic and task-oriented methods.
The findings also show the importance of semantic and syntactic information, and feasibility of automation and generalization with the promising audio-only and task-oriented methods for the AD detection task.

\end{abstract}
\begin{keywords}
Alzheimer’s disease, task-oriented, pretrained embeddings, transfer learning, multimodality
\end{keywords}

\section{Introduction}
\label{sec:into}

Alzheimer’s disease (AD) is a progressive neurodegenerative disease that causes gradual, irreversible deterioration in cognitive domains (memory, communication, etc.). Since there is no effective treatment for AD currently, early detection of this disease is particularly crucial for timely intervention and better disease control~\cite{mueller2005ways,rasmussen2019alzheimer}. Previous studies~\cite{appell1982study,cummings1988alzheimer,croot2000phonological,gayraud2011syntactic} have shown that symptoms of AD may be observable in spoken language at a very early stage, such as temporal disfluency, and difficulties in word finding and retrieval. These studies lay the theoretical foundation for using acoustic and linguistic information to screen for AD, which is attracting increasing interest from international research community.


The use of spoken language to screen for AD offers the advantages of affordability, accessibility and hence scalability, compared to conventional methods such as brain scans, blood tests and face-to-face neuropsychological assessments~\cite{gainotti2014neuropsychological}. Active research efforts are being devoted to finding spoken language features (both audio and linguistic features) as biomarkers of AD~\cite{fraser2016linguistic,weiner2019speech,pulido2020alzheimer,frankenberg2021verbal}.
For example, Weiner and Frankenberg \textit{et al.} \cite{weiner2019speech,frankenberg2021verbal} explored many traditional acoustic and linguistic features with a nested forward feature selection method, and found that the features on parts-of-speech, word categories and pauses are highly related to AD. Hence, possible approaches to design advanced algorithms to extract powerful acoustic and linguistic features to diagnose Alzheimer’s disease have become an emerging topic.


Inspired by the success of pretrained models, especially in speech (e.g., VGGish~\cite{koo2020exploiting}, Wav2Vec 2.0~\cite{balagopalan2021comparingAcoustic}, OpenL3~\cite{syed2021automated}) and text (e.g., BERT~\cite{balagopalan2020bert}, ERNIE~\cite{yuan2020disfluencies}, Glove~\cite{martinc2021temporal}), the development of AD detection is shifting from low-level features to higher-level representations in pretrained models. Although the BERT-like models have achieved promising performance on the AD detection task~\cite{yuan2020disfluencies,balagopalan2020bert,li2021comparative,wang2022exploring}, we note that these works mainly used higher-level representations of pretrained models, while features from intermediate layers may not have been devoted sufficient attention.
It is shown that intermediate layers encode rich hierarchical information for various features, e.g., surface features at the bottom, syntactic features in the middle and semantic features at the top~\cite{jawahar2019bert}.
Therefore, exploring and leveraging different levels of pretrained representations for better AD detection task is worthy of investigation.

The acoustic and linguistic features are typically used to distinguish the people with Alzheimer's disease from healthy controls, modeling the richness or disorder of participants' utterances to some extent. In addition to these features, it is also important to model the correctness and pertinence of participants' utterances for the cognitive tasks.
For example, in the widely used Cookie Theft Picture Description task, where people are asked to describe everything happening in a picture, Laura \textit{et al.} proposed information coverage measured by the statistics of the text and predefined referent~\cite{hernandez2018computer}.
This motivates us to propose features for modeling both the characteristics of disorder and the pertinence to the cognitive tasks.



In this work, we propose several efficient strategies to extract AD-related cues from embeddings of pretrained models, including aggregation along the layer and time dimension. We also use these extracted embeddings to measure task-related pertinence by correlation operation.
To validate the effectiveness of our proposed method, we conduct experiments on the Alzheimer’s Dementia Recognition through Spontaneous Speech (ADReSS) dataset \cite{luz2020alzheimer} with a binary classification setup. The proposed models are evaluated with accuracy and F1 scores, and obtain comparable results over other state-of-the-art methods in recent literature.
It's also exciting to find that the acoustic representations are comparable to the linguistic features, which could be more robust and generalizable in multilingual tasks and more helpful for fully automatic AD detection tasks.

The rest of the paper is organized as follows. Section \ref{sec:method} introduces the proposed methods for modeling with acoustic, linguistic and visual embeddings. Then, Section \ref{sec:exp} describes the experimental setups and results, as well as further analysis of the proposed methods and comparison with recent literature. Finally, Section \ref{sec:conclusion} concludes the paper and presents possible future research directions.

\section{Methodology}
\label{sec:method}
This section presents our methodology. Firstly, we describe the high-level embeddings for the audio and the text modalities. Then we introduce the measurement of task-related pertinence using these embeddings. And finally, we illustrate the feature extractor and classifier for the AD detection task.

\subsection{Acoustic \& Linguistic Embeddings}
To obtain rich characteristics for the AD detection, we adopt different pretrained models for speech and text to extract acoustic and linguistic features respectively. 

For the acoustic features, we compared several self-supervised (SSL) pretrained models (e.g., Wav2Vec 2.0~\cite{baevski2020wav2vec}, HuBERT~\cite{hsu2021hubert}, WavLM~\cite{chen2022wavlm}), and a weakly-supervised Whisper model~\cite{radford2022robust}.
The Wav2vec 2.0 is a model that jointly learns contextualized speech representations and an inventory of discretized speech units.
The HuBERT introduces a prior lexicon based on offline clustering to provide pseudo labels for speech units. The model is trained to predict the cluster assignments from the input speech units, which encourages the model to learn a combined acoustic and language model.
In order to solve full-stack downstream speech tasks, WavLM jointly learns masked speech prediction and denoising, by using some simulated noisy or overlapped
speech data \cite{chen2022wavlm}. The gated relative position bias is utilized to capture the sequence ordering of input speech better. With these improvements, WavLM is effective for not only the ASR task, but also several downstream tasks~\cite{horiguchi2020end}.

However, the speech of individuals with AD may differ from that of a general speaker, which may affect the performance of the self-supervised pretrained models on AD detection. A recent work, named Whisper~\cite{radford2022robust}, has achieved state-of-the-art performance on many ASR tasks, which is trained on a web-scale 680,000 hours in a weakly supervised multilingual multitask fashion. Therefore, compared with previous models, Whisper is more advantageous and robust for application to different downstream tasks. Considering these merits of Whisper in processing speech tasks, we adopt it as the default acoustic feature extractor. For linguistic features, we choose BERT~\cite{devlin2018bert}, a Transformer-based~\cite{vaswani2017attention} pretrained model, as the basic backbone network, which adopts masked language modeling and next sentence prediction as the pre-training objective. 

Whisper calculates the logarithm Mel-Spectrum and then connects with two Convolution layers, followed by the Transformer layers. We feed audio into Whisper and use its produced features as acoustic representation. So for each second of audio, we can obtain features from different Transformer layers, which have a dimension of $50 \times H_a \times L_a$ (dimensions of time, feature and layer respectively). And for the text modality, we feed the transcribed text from the audios of patients into BERT model to obtain the features as $1 \times H_t \times L_t$ of each token.

\subsection{Task-related Information}
\begin{figure}[htb]
\centering
\includegraphics[width=0.9\linewidth]{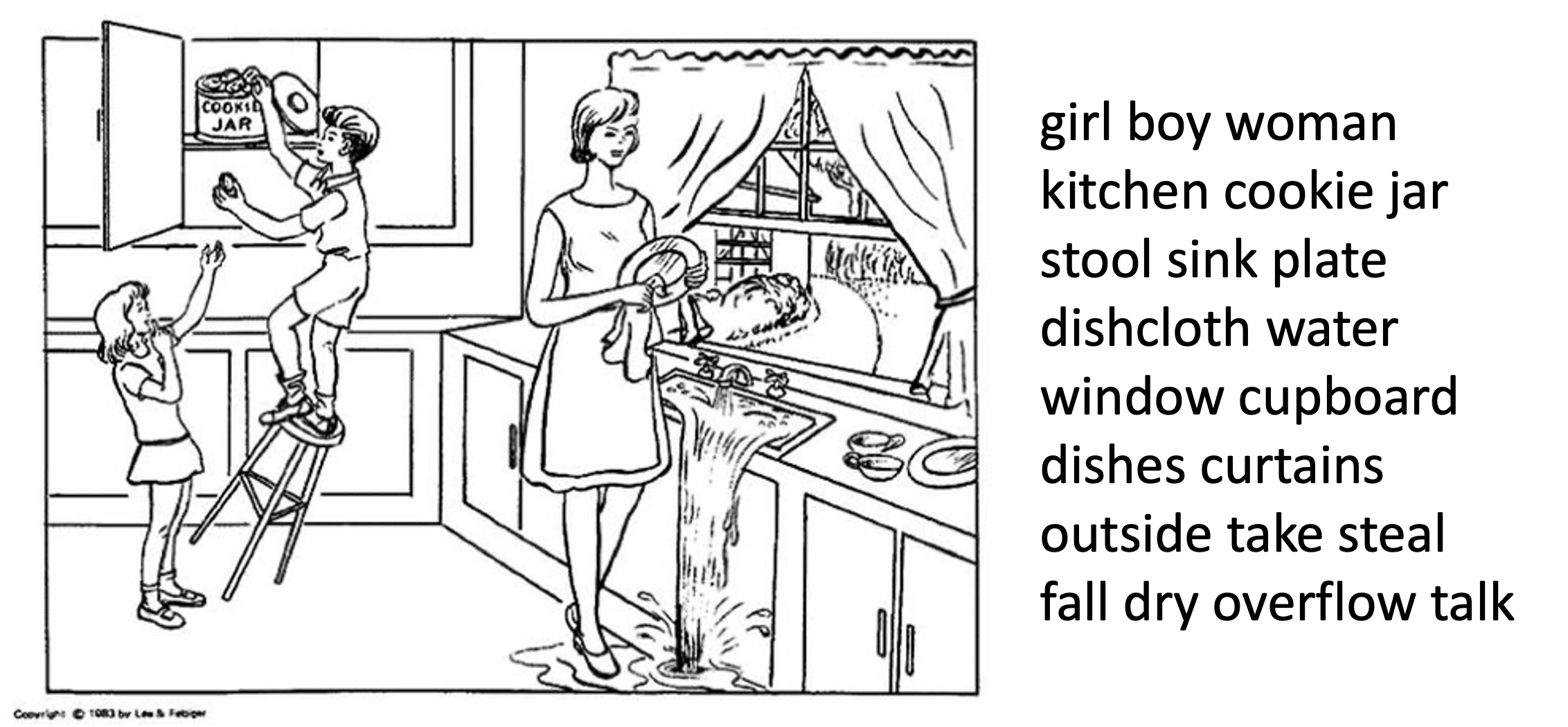}
\caption{The Cookie Theft Picture and pre-defined keywords.}
\label{fig:cookie}
\end{figure}

For the Cookie Theft Picture Description Task, we aim to integrate information from the picture using a series of pre-defined keywords, as shown in Fig.~\ref{fig:cookie}. The keywords include the named entities (nouns) and actions (verbs) happening in the picture.
We calculate the correlation between the linguistic embeddings of the spoken utterances (describing the picture) and the pre-defined keywords for the picture.
Formally, let the extracted embeddings of spoken utterances and task-related keywords be $z_u$ and $z_k$, then the task-related correlation is $Corr. = z_u \times z_k$ by element-wise production.

\subsection{Feature Extractor}

\begin{figure}[htb]
\centering
\includegraphics[width=0.7\linewidth]{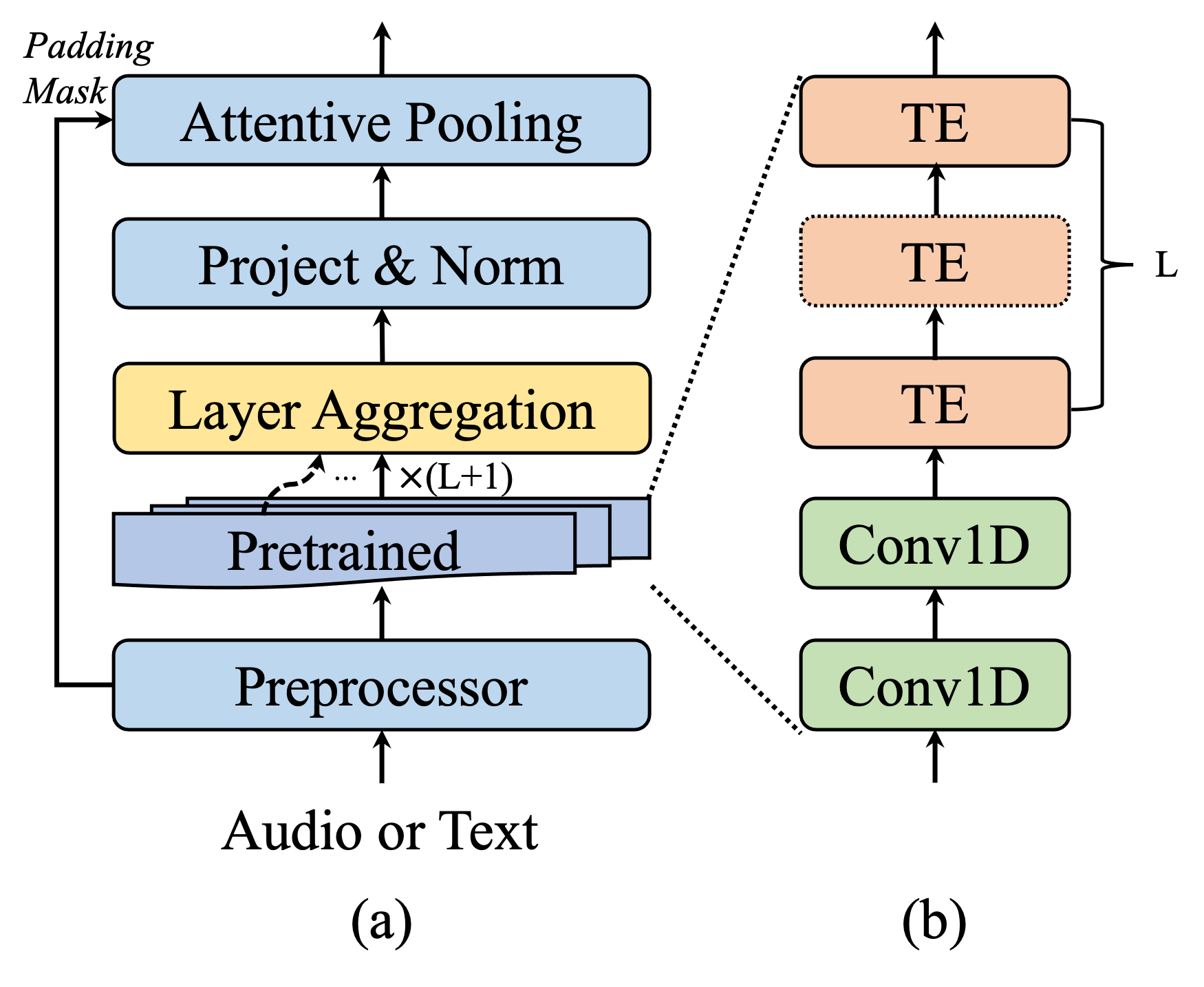}
\vspace{-1em}
\caption{Feature extraction with pretrained models. ``TE'' denotes the Transformer Encoder, ``L'' denotes the number of TE layers. Compared to BERT, Whisper has two Conv1D before the stack of TEs.}
\label{fig:feat}
\end{figure}

As mentioned earlier, the raw embeddings extracted from the Whisper or BERT for the two modalities are large and redundant. To further condense the information which should be helpful for the AD detection task, we propose several subsequent modules as illustrated in Fig.~\ref{fig:feat}.
First, we perform layer aggregation to calibrate layer-wise feature responses. Here, we adopt two strategies: the first is ``weighted sum (WS)'', which sums up all elements with learnable weights; the second is ``maximum single (MS)'', which selects one single layer with the best performance.
Second, we project the calibrated features into a lower dimensionality to reduce feature redundancy while retaining intra-class variability. The projector is an MLP with Layer-Normalization, which can also map audio and text modalities into the same dimensional space for fusion.
Third, we use an attentive temporal pooling layer~\cite{santos2016attentive} to compress the sequence with variant time lengths into a fixed-length vector, and capture richer statistics of the temporal features for the AD detection task.
Finally, we average the features from different segments for each speaker.

\subsection{Classifier}

The extracted features are either fed into a classifier directly or combined with others with concatenation operation. The classifier used in this work is a fully-connected layer that produces the probabilities of binary classification between an individual having AD or being a Healthy Control (HC).

\section{Experiments}
\label{sec:exp}
In this section, we describe the corpus, experimental setup and results.

\subsection{Corpus}
The corpus used in this work comes from the Alzheimer’s Dementia Recognition Through Spontaneous Speech (ADReSS) Challenge 2020 corpus~\cite{luz2020alzheimer}. This challenge selects a sub-task of Pitt Corpus in the DementiaBank database~\cite{becker1994natural}, which requires all the participants to describe the Cookie Theft picture as shown in Fig.~\ref{fig:cookie}. The ADReSS corpus consists of 156 different English speakers' audio samples with corresponding transcripts. Among them, 78 of the speakers are healthy control (35 male, 43 female) while the rest are with AD (35 male, 43 female). The corpus is divided into a standard train (108 speakers, about 2 hours) and test (48 speakers, about 1 hour) sets with balanced distributions of age, gender and disease conditions.

\subsection{Experimental Setup}

\subsubsection{Data Preprocessing}
At the beginning of the experiments, we preprocess the speech with enhancement~\cite{hao2021fullsubnet} and normalization methods for internal consistency of the data. In addition, we use data augmentation to enrich the data and improve the robustness by slightly changing the acoustic characteristics with minor distortions. Specifically, we used three strategies, pitch-shifting, speed-perturbation, and dithering for each input waveform during the training stage~\cite{cariani1996neural,colosi1998efficient,schuchman1964dither}.
The shifted ranges of pitch and speed are [-100, 100] semitones and [-0.05, +0.05] rates, respectively.

Notably, the data on AD investigated in this work are in long-form, e.g., several minutes of spontaneous speech associated with transcripts from the picture description task. However, the inputs of the high-level pretrained models are usually within 30 seconds for audio or 512 tokens for text. To deal with this issue, we slice the experimental audio into 30-second segments with a hop ratio of 0.25, and obtain the aligned transcripts that are less than 512 tokens. We then aggregate the extracted segment-level features for each speaker.

\subsubsection{Model \& Training Details}
The variants of the Whisper and BERT models adopted in this work are respectively the small and base-uncased versions pretrained on English corpora , both of which output 768-dimensional embeddings.
The projector is composed of two stacked 8-dimensional linear layers with layer normalization, and the classifier is a fully-connected linear classifier. We insert a dropout layer with a rate of 0.25 between the projector and temporal pooling layer.

The training loss of this work is set to be the binary cross-entropy loss. We use AdamW~\cite{loshchilov2017decoupled} as our optimizer with a learning rate of $1e-4$ and a weight decay of $1e-5$. The models are trained with a batch size of 16 for 50 epochs.

\subsubsection{Evaluation Protocols}

We evaluate the model performance on the unseen ADReSS test data, with the metrics of classification accuracy and macro F1 scores, the mean of class-wise F1-scores, that averaged on five random runs.
To better compare our proposed methods with previous literature, we choose multiple baselines, including Luz \textit{et al.} which used ComParE and Linguistic feature sets~\cite{luz2020alzheimer}, 
Koo~\cite{koo2020exploiting} and Syed \textit{et al.}~\cite{syed2021automated} which used acoustic pretrained models, and Yuan~\cite{yuan2020disfluencies}, Matej~\cite{martinc2021temporal} and Yi \textit{et al.}~\cite{wang2022exploring} which used linguistic pretrained models.

\subsection{Results}

\subsubsection{Results using Extracted Embeddings}

We compare the performance of acoustic and linguistic features with different aggregation strategies on the layer and time dimensions, including weighted-sum (WS) or single selected (Top, MS) for the layers, and mean (Mean) or attentive (Attention) pooling for the time-axis, as listed in the Table.~\ref{tab:feat}.

\begin{table}[htb]
\centering
\setlength\tabcolsep{1mm}
\resizebox{\linewidth}{!}{
\begin{tabular}{ l c c c c }
\toprule
Feature & Layer AGG & Time AGG & Accuracy(\%) & F1-score(\%) \\
\midrule
ComParE~\cite{luz2020alzheimer} & - & Mean & 62 & 62 \\
Linguistics~\cite{luz2020alzheimer} & - & Mean & 75 & 71 \\
\midrule
VGGish~\cite{koo2020exploiting} & Top & Mean & 72.92 & 72.62 \\
OpenL3~\cite{syed2021automated}$^{\star}$ & Top & Mean & 81.25 & 81.20 \\
ERNIE~\cite{yuan2020disfluencies} & Top & Mean & 85.4 & 85.3 \\
Glove~\cite{martinc2021temporal} & Top & Mean & 89.6 & - \\
BERT+Roberta~\cite{wang2022exploring}$^{\star}$ & Top & Mean & {\bf 91.7} & {\bf 91.7} \\
\midrule
Wav2vec 2.0 & WS & Mean & 77.50 & 76.69 \\
HuBERT & WS & Mean & 78.88 & 78.79 \\
WavLM & WS & Mean & 79.74 & 79.66 \\
Whisper & WS & Mean & 79.31 & 79.30 \\
BERT & WS & Mean & 87.50 & 87.47 \\

\midrule
WavLM & WS & Attention & 82.33 & 82.33 \\ 
Whisper & WS & Attention & 81.47 & 81.46 \\
BERT & WS & Attention & 88.79 & 88.76 \\
\midrule
WavLM & MS & Attention & 85.78 & 85.78 \\
Whisper & MS & Attention & \textbf{88.79} & \textbf{88.79} \\
BERT & MS & Attention & \textbf{90.09} & \textbf{90.07} \\
\bottomrule
\end{tabular}
}
\caption{Performance based on acoustic features or representations. ``AGG'' denotes ``aggregation'', ``WS'' denotes ``weighted sum'', and ``MS'' denotes ``maximum single''. ``$\star$'' denotes ensemble systems.}
\label{tab:feat}
\vspace{-1em}
\end{table}

It can be observed that, (i) the best single selected method outperforms the weighted-sum method for layer-wise representations; (ii) the attentive pooling method outperforms the mean pooling for the time-axis information; (iii) Whisper outperforms other acoustic models with the MS and attentive pooling methods; and (iv) linguistic representations generally outperform the acoustic representations.

\begin{figure}[htb]
\centering
\vspace{-1em}
\centerline{\includegraphics[width=0.9\linewidth]{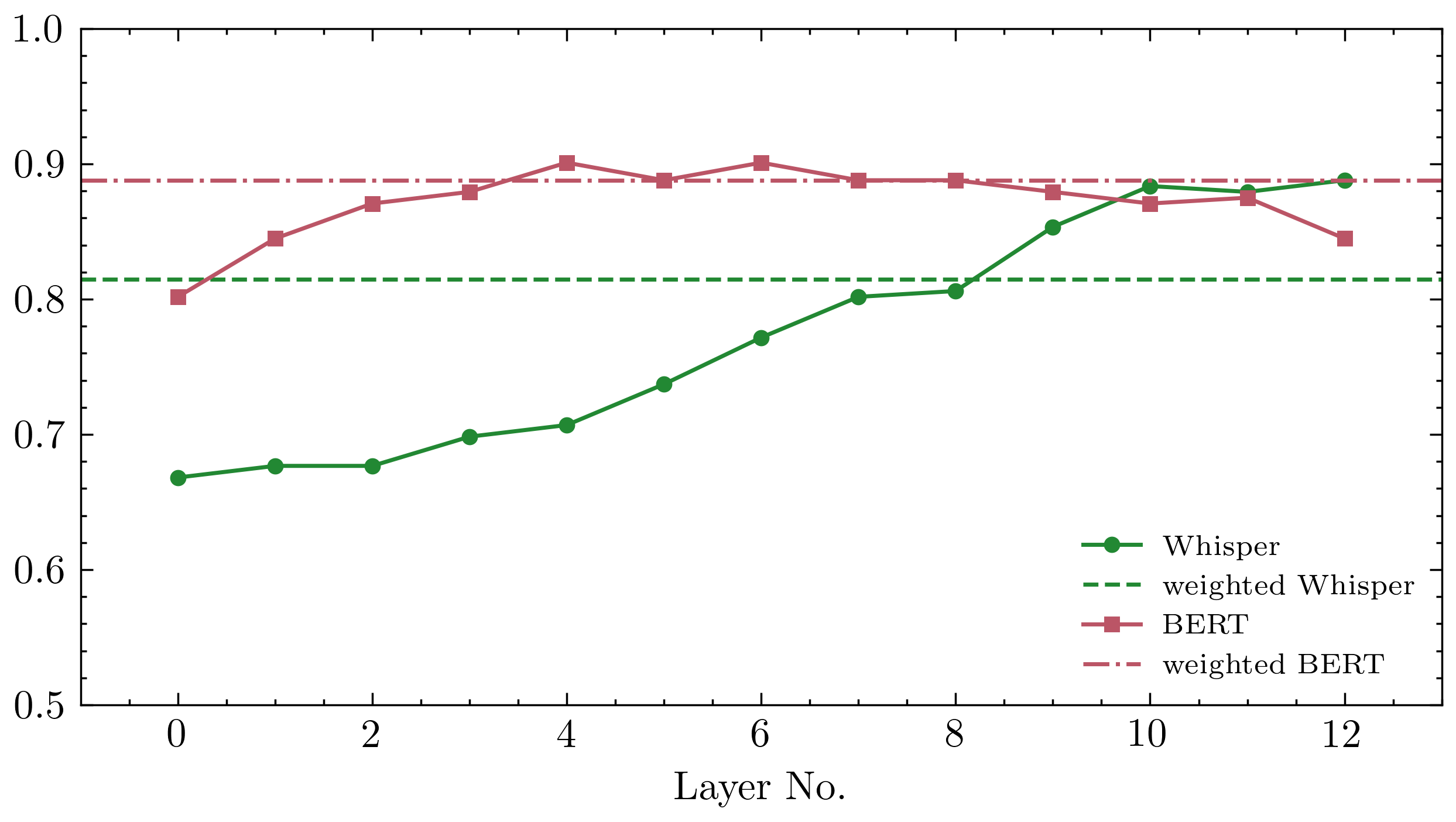}}
\vspace{-1em}
\caption{Effectiveness of various layers of pretrained Whisper and BERT. The solid and dashed lines denote systems using only one single layer or a weighted sum of all layers respectively.}
\label{fig:layer}
\end{figure}

For the first observation, we find that the weights in the WS method are not well-matched with the performance distributions using a single layer. For example, as shown in Fig.~\ref{fig:layer}, the topmost layers (larger layer No.) of Whisper and the middle layers of BERT show higher performance for the AD detection task, but the learned weights may also be distracted by the bottom layers in the WS method, which harms the performance. We study the performance on each layer in depth.
For the acoustic models, the observation of that the topmost layers outperforms others coincides with previous research that higher layers capture more word and semantic information \cite{pasad2021layer}, which are crucial for AD detection. This also supports the way of using the topmost layer for AD detection that is widely adopted in previous research \cite{koo2020exploiting,balagopalan2021comparing,syed2021automated}.
For the linguistic models, the observation of that middle layers outperforms others implies that the syntactic information is more important than the semantic information \cite{jawahar2019bert} for the AD detection, which is also intuitive, since the syntactic information can model the cognitive disorder better. 

The second observation supports that the attentive pooling method captures richer statistics of temporal features than the mean pooling method.
The first and second observations also show that the layer and time dimensions of pretrained models have different importance in the AD detection task. Take the Whisper model as an example, the Attention-based time aggregation improves the accuracy scores by about 2.7\% relatively, while the MS layer aggregation is more effective and further improves it by about 9\% relatively.

The third observation reflects the robustness and effectiveness of the Whisper model for AD detection.
And the forth observation coincides with previous research \cite{pulido2020alzheimer,li2021comparative}.
It is also interesting to find that the performance of acoustic models is now comparable with that of linguistic ones, which worse a lot than the latter in the past.
The underlying mechanism of pretrained encoders is difficult to interpret, but we could intuitively explain the finding in terms of representation and pathology.
On the one hand, the high-level pretrained acoustic encoders could extract the semantic 
information, especially from the top layers, that is similar to the linguistic features and helpful for the AD detection task.
On the other hand, AD affects the participants' phonology and articulation~\cite{croot2000phonological,gayraud2011syntactic}, such as dysfluencies (aphasic to some degree) and hesitations, while acoustic encoders could also extract these features that may not be easily extracted from the text.
These encoded semantic and acoustic features could make acoustic models comparable to linguistic methods in the AD detection task.
The promising performance of acoustic models not only promotes the fully-automation of the AD detection task, but also could be helpful for multilingual generalization since some acoustic characteristics are more ``universal" across languages than linguistic ones.

\subsubsection{Results using Task-related Information}
\begin{table}[htb]
\centering
\setlength\tabcolsep{1mm}
\begin{tabular}{ l c c }
\toprule
Type & Accuracy(\%) & F1-score(\%) \\
\midrule
None & 52.34 & 34.36 \\
Nouns & 85.16 & 85.11 \\
Verbs & 83.59 & 83.58 \\
Nouns + Verbs & {\bf 85.94} & {\bf 85.88}\\
\bottomrule
\end{tabular}
\caption{Results using correlation features between text and keywords of ``None'' (empty), ``Nouns'', ``Verbs'' and both.}
\label{tab:key}
\vspace{-1em}
\end{table}




We also compare the task-related correlation features with different keyword lists, including ``None'' (empty), ``Nouns''-only (named entities), ``Verbs''-only (actions) and combination of ``Nouns'' and ``Verbs''. It can be found that using keywords of ``Nouns`` performs better than using that of ``Verbs'' for AD detection task, which implies the ability of named entities retrieval are affected by Alzheimer's Disease and coincides with the fact that Named Task is important in the clinical cognitive tasks.
We also decouple the effect of task-related correlation by using an empty keyword list, and the performance drops to 52.34\%, which can be view as a random guess.
It can be also observed that using correlation with task-related keywords can achieve over 80\% accuracy scores, which is better than the linguistic measures in \cite{luz2020alzheimer}.

\subsubsection{Feature Combination Results}

Finally, we compare the performance of a combination of the acoustic (Whisper), linguistic (BERT) and task-related correlation (Corr.) features, as listed in Table~\ref{tab:mm}.
Generally, the combination of different modalities outperforms the other systems with a superior performance of 91.41\% accuracy, which implies that complementary information from various modalities helps AD detection.

\begin{table}[htb]
\centering
\setlength\tabcolsep{1mm}
\begin{tabular}{ l c c }
\toprule
Feature & Accuracy(\%) & F1-score(\%) \\
\midrule
OpenL3 + Roberta~\cite{syed2021automated}$^{\star}$ & 89.85 & 89.54 \\
Temporal + Glove~\cite{martinc2021temporal} & 91.67 & - \\
\midrule
Whisper & 88.79 & 88.79 \\
BERT & 90.09 & 90.07 \\
Corr. & 85.94 & 85.88 \\
Whisper + BERT & 91.19 & 91.19 \\
Whisper + Corr. & 89.84 & 89.83 \\
BERT + Corr. & 90.62 & 90.60 \\
Whisper + BERT + Corr. & \textbf{91.41} & \textbf{91.38} \\
\bottomrule
\end{tabular}
\caption{Results on a combination of features. ``Corr.'' denotes correlation embeddings between utterance and keywords from picture. ``$\star$'' denotes ensemble systems.}
\label{tab:mm}
\vspace{-1em}
\end{table}

\section{Conclusion}
\label{sec:conclusion}

In this work, we explored pretrained representations from different modalities for Alzheimer's Disease detection.
Experiments on the ADReSS corpus have shown superior performance by using acoustic and linguistic embeddings, as well as task-related keywords.
Results indicate that the top layers of pretrained acoustic models and the middle layers of pretrained linguistic models provide features that are more important for the AD detection task.
It also shows that the correlation and richness measured by task-related keywords and described utterances could also help the AD detection task.

Future work will include multimodal pretrained embeddings to model visual-textual relationships, more efficient fusion strategies to boost performance, as well as using acoustic embeddings for automatic and multilingual AD detection tasks.

\section{Acknowledgements}
\label{sec:ack}
This project is partially supported by the HKSARG Research Grants Council's Theme-based Research Grant Scheme (Project No. T45-407/19N).

\bibliographystyle{IEEEbib}
\bibliography{main}

\begin{thebibliography}{10}

\bibitem{mueller2005ways}
S.~G. Mueller, M.~W. Weiner, et~al.,
\newblock ``Ways toward an early diagnosis in alzheimer’s disease: the
  alzheimer’s disease neuroimaging initiative (adni),''
\newblock {\em Alzheimer's \& Dementia}, 2005.

\bibitem{rasmussen2019alzheimer}
J. Rasmussen and H. Langerman,
\newblock ``Alzheimer’s disease--why we need early diagnosis,''
\newblock {\em Degenerative neurological and neuromuscular disease}, 2019.

\bibitem{appell1982study}
J. Appell, A. Kertesz, and M. Fisman,
\newblock ``A study of language functioning in alzheimer patients,''
\newblock {\em Brain and language}, 1982.

\bibitem{cummings1988alzheimer}
J.~L. Cummings, A. Darkins, et~al.,
\newblock ``Alzheimer's disease and parkinson's disease: comparison of speech
  and language alterations,''
\newblock {\em Neurology}, 1988.

\bibitem{croot2000phonological}
K. Croot, J.~R. Hodges, et~al.,
\newblock ``Phonological and articulatory impairment in alzheimer's disease: a
  case series,''
\newblock {\em Brain and language}, 2000.

\bibitem{gayraud2011syntactic}
F. Gayraud, H.-R. Lee, and M. Barkat-Defradas,
\newblock ``Syntactic and lexical context of pauses and hesitations in the
  discourse of alzheimer patients and healthy elderly subjects,''
\newblock {\em Clinical linguistics \& phonetics}, 2011.

\bibitem{gainotti2014neuropsychological}
G. Gainotti, D. Quaranta, et~al.,
\newblock ``Neuropsychological predictors of conversion from mild cognitive
  impairment to alzheimer's disease,''
\newblock {\em Journal of Alzheimer's disease}, 2014.

\bibitem{fraser2016linguistic}
K.~C. Fraser, J.~A. Meltzer, and F. Rudzicz,
\newblock ``Linguistic features identify alzheimer’s disease in narrative
  speech,''
\newblock {\em Journal of Alzheimer's Disease}, 2016.

\bibitem{weiner2019speech}
J. Weiner, C. Frankenberg, et~al.,
\newblock ``Speech reveals future risk of developing dementia: Predictive
  dementia screening from biographic interviews,''
\newblock in {\em ASRU}. IEEE, 2019.

\bibitem{pulido2020alzheimer}
M.~L.~B. Pulido, J.~B.~A. Hern{\'a}ndez, et~al.,
\newblock ``Alzheimer's disease and automatic speech analysis: a review,''
\newblock {\em Expert systems with applications}, 2020.

\bibitem{frankenberg2021verbal}
C. Frankenberg, J. Weiner, et~al.,
\newblock ``Verbal fluency in normal aging and cognitive decline: Results of a
  longitudinal study,''
\newblock {\em Computer Speech \& Language}, 2021.

\bibitem{koo2020exploiting}
J. Koo, J.~H. Lee, et~al.,
\newblock ``Exploiting multi-modal features from pre-trained networks for
  alzheimer's dementia recognition,''
\newblock in {\em INTERSPEECH}, 2020.

\bibitem{balagopalan2021comparingAcoustic}
A. Balagopalan and J. Novikova,
\newblock ``Comparing acoustic-based approaches for alzheimer's disease
  detection,''
\newblock {\em arXiv preprint arXiv:2106.01555}, 2021.

\bibitem{syed2021automated}
Z.~S. Syed, M.~S.~S. Syed, et~al.,
\newblock ``Automated recognition of alzheimer’s dementia using
  bag-of-deep-features and model ensembling,''
\newblock {\em IEEE Access}, 2021.

\bibitem{balagopalan2020bert}
A. Balagopalan, B. Eyre, et~al.,
\newblock ``To bert or not to bert: comparing speech and language-based
  approaches for alzheimer's disease detection,''
\newblock {\em arXiv preprint arXiv:2008.01551}, 2020.

\bibitem{yuan2020disfluencies}
J. Yuan, Y. Bian, et~al.,
\newblock ``Disfluencies and fine-tuning pre-trained language models for
  detection of alzheimer's disease.,''
\newblock in {\em INTERSPEECH}, 2020.

\bibitem{martinc2021temporal}
M. Martinc, F. Haider, et~al.,
\newblock ``Temporal integration of text transcripts and acoustic features for
  alzheimer's diagnosis based on spontaneous speech,''
\newblock {\em Frontiers in Aging Neuroscience}, 2021.

\bibitem{li2021comparative}
J. Li, J. Yu, et~al.,
\newblock ``A comparative study of acoustic and linguistic features
  classification for alzheimer's disease detection,''
\newblock in {\em ICASSP}. IEEE, 2021.

\bibitem{wang2022exploring}
Y. Wang, T. Wang, et~al.,
\newblock ``Exploring linguistic feature and model combination for speech
  recognition based automatic ad detection,''
\newblock {\em INTERSPEECH}, 2022.

\bibitem{jawahar2019bert}
G. Jawahar, B. Sagot, and D. Seddah,
\newblock ``What does {BERT} learn about the structure of language?,''
\newblock in {\em ACL}, 2019.

\bibitem{hernandez2018computer}
L. Hern{\'a}ndez-Dom{\'\i}nguez, S. Ratt{\'e}, et~al.,
\newblock ``Computer-based evaluation of alzheimer’s disease and mild
  cognitive impairment patients during a picture description task,''
\newblock {\em Alzheimer's \& Dementia: DADM}, 2018.

\bibitem{luz2020alzheimer}
S. Luz, F. Haider, et~al.,
\newblock ``{Alzheimer’s Dementia Recognition Through Spontaneous Speech: The
  ADReSS Challenge},''
\newblock {\em INTERSPEECH}, 2020.

\bibitem{baevski2020wav2vec}
A. Baevski, Y. Zhou, et~al.,
\newblock ``wav2vec 2.0: A framework for self-supervised learning of speech
  representations,''
\newblock {\em NeurIPS}, 2020.

\bibitem{hsu2021hubert}
W.-N. Hsu, B. Bolte, et~al.,
\newblock ``Hubert: Self-supervised speech representation learning by masked
  prediction of hidden units,''
\newblock {\em IEEE/ACM TASLP}, 2021.

\bibitem{chen2022wavlm}
S. Chen, C. Wang, et~al.,
\newblock ``Wavlm: Large-scale self-supervised pre-training for full stack
  speech processing,''
\newblock {\em IEEE J. Sel. Top. in Signal Process.}, 2022.

\bibitem{radford2022robust}
A. Radford, J.~W. Kim, et~al.,
\newblock ``Robust speech recognition via large-scale weak supervision,''
\newblock Tech. {R}ep., OpenAI, 2022.

\bibitem{horiguchi2020end}
S. Horiguchi, Y. Fujita, et~al.,
\newblock ``End-to-end speaker diarization for an unknown number of speakers
  with encoder-decoder based attractors,''
\newblock {\em INTERSPEECH}, 2020.

\bibitem{devlin2018bert}
J. Devlin, M.-W. Chang, et~al.,
\newblock ``Bert: Pre-training of deep bidirectional transformers for language
  understanding,''
\newblock {\em arXiv preprint arXiv:1810.04805}, 2018.

\bibitem{vaswani2017attention}
A. Vaswani, N. Shazeer, et~al.,
\newblock ``Attention is all you need,''
\newblock {\em NeurIPS}, 2017.

\bibitem{santos2016attentive}
C.~d. Santos, M. Tan, et~al.,
\newblock ``Attentive pooling networks,''
\newblock {\em arXiv preprint arXiv:1602.03609}, 2016.

\bibitem{becker1994natural}
J.~T. Becker, F. Boiler, et~al.,
\newblock ``The natural history of alzheimer's disease: description of study
  cohort and accuracy of diagnosis,''
\newblock {\em Archives of neurology}, 1994.

\bibitem{hao2021fullsubnet}
X. Hao, X. Su, et~al.,
\newblock ``Fullsubnet: A full-band and sub-band fusion model for real-time
  single-channel speech enhancement,''
\newblock in {\em ICASSP}. IEEE, 2021.

\bibitem{cariani1996neural}
P.~A. Cariani and B. Delgutte,
\newblock ``Neural correlates of the pitch of complex tones. ii. pitch shift,
  pitch ambiguity, phase invariance, pitch circularity, rate pitch, and the
  dominance region for pitch,''
\newblock {\em Journal of neurophysiology}, 1996.

\bibitem{colosi1998efficient}
J.~A. Colosi and M.~G. Brown,
\newblock ``Efficient numerical simulation of stochastic internal-wave-induced
  sound-speed perturbation fields,''
\newblock {\em The Journal of the Acoustical Society of America}, 1998.

\bibitem{schuchman1964dither}
L. Schuchman,
\newblock ``Dither signals and their effect on quantization noise,''
\newblock {\em IEEE TCT}, 1964.

\bibitem{loshchilov2017decoupled}
I. Loshchilov and F. Hutter,
\newblock ``Decoupled weight decay regularization,''
\newblock {\em arXiv preprint arXiv:1711.05101}, 2017.

\bibitem{pasad2021layer}
A. Pasad, J.-C. Chou, and K. Livescu,
\newblock ``Layer-wise analysis of a self-supervised speech representation
  model,''
\newblock in {\em ASRU}. IEEE, 2021.

\bibitem{balagopalan2021comparing}
A. Balagopalan and J. Novikova,
\newblock ``Comparing acoustic-based approaches for alzheimer's disease
  detection,''
\newblock {\em arXiv preprint arXiv:2106.01555}, 2021.

\end{thebibliography}

\end{document}